\documentstyle[aps,preprint,prl,epsfig]{revtex}

\draft
\newcommand{\ba}{\begin{eqnarray}}
\newcommand{\ea}{\end{eqnarray}}
\newcommand{\bmath}{\begin{mathletters}}
\newcommand{\emath}{\end{mathletters}}
\newcommand{\ban}{\begin{eqnarray*}}
\newcommand{\ean}{\end{eqnarray*}}

\begin{document}



\title{General pairing interactions and pair truncation approximations
for fermions in a single-$j$ shell }
\author{ Y. M. Zhao$^{a,b}$,
      A. Arima$^{c}$,  J. N. Ginocchio$^{d}$, and N. Yoshinaga $^a$ }

\address{$^a$ Department of Physics, Saitama University,
Saitama-shi, Saitama 338 Japan }
\address{$^b$ Department of Physics,  Southeast University, Nanjing
210018 China }
\address{$^c$ The House of Councilors, 2-1-1 Nagatacho,
Chiyodaku, Tokyo 100-8962, Japan }
\address{$^d$ MS B283, Los Alamos National Laboratory,
Los Alamos, NM 87545, USA}



\maketitle

\begin{abstract}
We investigate Hamiltonians with attractive interactions between
pairs of fermions
coupled to angular momentum $J$.
We show that  pairs with spin $J$  are
reasonable building blocks for the low-lying states. For systems with
only a $J =
J_{max}$ pairing interaction, eigenvalues  are found to be approximately
integers  for a large array of states, in particular
for those with total angular momenta  $I \le 2j$.
For $I=0$ eigenstates of four fermions
in a single-$j$ shell we show that there is only one
non-zero eigenvalue. We address these observations
using the nucleon pair approximation of the shell model
and relate our results  with a number of
currently interesting problems.
\end{abstract}



\newpage
\section {Introduction}
Since pairing has proven to be important in atomic, nuclear, and
condensed matter physics, pair truncation approximations to many body
wavefunctions have been extensively studied. The first example is the
seniority scheme, introduced by Racah
\cite{Racah,Talmi}, for the classification of states  in atomic spectra
and later  applied extensively in nuclear physics, where $S$ pairs
with angular momentum zero are related with a strong and attractive monopole
pairing interaction. The second example is
the  interacting boson model  (IBM)
introduced by Arima and Iachello  \cite{Arima}, where
the low-lying excitations of complex even-even nuclei
are described successfully by $s$ bosons which
correspond to  correlated  $S$ nucleon pairs
with angular momentum zero  and $d$ bosons
which correspond  to  $D$ nucleon pairs with angular momentum two. Again,
the   success of the IBM in nuclear physics partly comes from the
validity of the pairing plus quadrupole-quadrupole force for
effective interactions between valence nucleons. Monopole and quadrupole
pairing are important as well in low and high
temperature superconductiviy in materials \cite{muller,iachello}.

In this paper we investigate the general pair truncation
approximation
for fermions in a single-$j$ shell.
The examples explored may provide a clue
as how to classify the states which come from the diagonalization
of an attractive pairing interaction for which two fermions are
coupled to an angular momentum $J$,
\begin{eqnarray}
&&  H_J =
-\sum_{M = -J}^J A_M^{J \dagger} \ A_{M}^{J }
\nonumber \\
&&  A_M^{J \dagger} = \frac{1}{\sqrt{2}} \left[ a_{j}^{\dagger}
\times a_{j}^{\dagger}
     \right]^J, ~ ~
     {A}_M^J = - (-1)^M\frac{1}{\sqrt{2}} \left[ \tilde{a}_{j} \times
     \tilde{a}_{j} \right]_{-M}^J.   \label{pair}
\label{H}
\end{eqnarray}
where $[~]_M^J$ means coupled to angular momentum $J$ and projection $M$.
Most of examples pursued in this paper have $n$ = 4, where  $n$ is the
nucleon number.

We first point out in
this paper that the low-lying eigenstates of (\ref{H}) can be
approximated by wavefunctions with pairs with angular momentum $J$ only.
We shall next show that a large array of eigenvalues
of four nucleons in a single-$j$ shell are
asymptotic integers when $J \sim J_{max} = 2j-1$ and the total angular
momentum, $I$, is not
very close to
$I_{max}= 4j-6$,
    and that this phenomenon  originates from validity
of the pair truncation scheme and special features of
coupling coefficients.
We shall finally prove that the pair Hamiltonian
(\ref{H}) has exactly {\it one
and only one} non-zero eigenvalue for  four fermion eigenstates with
total angular
momentum zero,
$I=0$. This sheds light  on the problem of angular momentum zero
ground state dominance
in  many-body systems interacting by random interactions \cite{Johnson}.

\section{Comparison of the Pair Approximation to the Shell Model}

Fig. 1 compares the exact ground state angular momentum $I$
  for four nucleons in a
single-$j$ shell interacting by the attractive pair Hamiltonian
$H_{J}$ with the
angular momentum $I$ of the ground state in the truncated space of
pairs coupled to
angular momenutm $J$ only. We have examined all the cases up to $J=20$ and
$j \le 31/2$ but here we show only two  typical examples with $J=6$ and 14.
    In the case of $J=0$, the
    seniority scheme ($S$-pair approximation) produces the
exact ground state.
In  case of $J=2$ and  $n=4$,  a $D$-pair
approximation is found to be always very good.
When $J > 2$, the $J$-pair approximation
of low-lying states is not perfect but always
very reasonable. In Fig. 1,   most of  ground state angular momenta
are correctly reproduced by the $J$-pair truncation.
In Fig. 1b), there are two exceptions in which the
ground state is not correctly given by two $J=14$ pairs:
$2j=19$ and  25.

Even in those cases where the ground state angular momenta
are  not correctly predicted by the $J$-pair truncation,
the low-lying state energies are   reasonably
reproduced (including the binding energies).
As a ``bad" example in which  the ground
state angular momentum is not reproduced, we show
in Fig. 2 the case of four nucleons in a
$j=25/2$ shell with $J=14$.
The calculated
levels using two $J=14$ pairs are shown in the first column. The next
two columns
are the shell model states obtained by diagonalizing the Hamiltonian
in the full space. The states in the second column are the shell model
states correponding to the pair truncation states in the first
column. All the levels
below $0^+$ are included.
One sees that the lowest four states
$2_1^+$, $6_1^+$, $12_1^+$, and $10_1^+$
are well approximated by two $J=14$ pairs, although
the ground state angular momentum is not correctly given by this $J=14$ pair
approximation.
The $0_1^+$ state is always precisely
reproduced  (as we shall see  there is
only one non-zero eigenvalue for $I=0$ states
and that eigenstate is constructed by $J$ pairs).
The states with odd $I$
are always outside the pair truncation space but their energies are
quite high in all cases that we have checked. In Fig. 2, the $9_1^+$ is
the lowest state with odd value of $I$. Below the $9_1^+$ there are
10 states with even values of $I$ and most of which
are nicely described  by the $J=14$ pair approximation.
The angular momentum of those states
for which excited energies are not
well reproduced by two $J$ pairs
are labled by italic font in Fig. 2.

For $J=J_{max}= 2j-1$, the $I=I_{max}= 4j-6$ or $I=I_{max}-2$ states
are always the lowest.
These two states may be constructed by  pairs with angular momentum
either $J_{max}$
or $J_{max}-2$. However,  pairs with angular momentum $J_{max}-2$  do not
present a good classification for other $I$ states while
those with angular momentum $J_{max}$ do. 

For n=3, the $J$-pair truncation describes the low-lying
states precisely. For $n$=5 and 6 cases that we have
examined, the low-lying states are reasonably approximated
by the $J$-pair truncation. We note without 
details that bosons with spin $l$ exhibit a similar
situation. It would be interesting to know the situation
in more complicated systems.

\section {Integer Eigenvalues}
We next report a very interesting regularity in the spectrum of the
Hamiltonian (\ref{H}) with $J=J_{max}=2j-1$.
The eigenvalues of most states with $I \le 2j-3$  are
found to be very close to
integers corresponding to the number of  pairs  with
angular momentum $J_{max}$ except for a very few eigenvalues.
Taking four nucleons  in a $j=31/2$  shell as an example, the
diagonalization of $H_J$ ($J=30$) gives
``integer" eigenvalues for  low $I$ states$--$
all eigenvalues are 0, $- 1$, and $- 2$ to within a precision of
0.01 for {\bf all} states with $I < 22$.
For states with $22 \le I \le 52$, these
three ``integer"  eigenvalues continue to be valid except that
7 eigenvalues  which are not close to 0, $ -1$ or $ - 2$
come in. These ``non-integer" eigenvalues are very stable in magnitude
for states with $22\le I \le 52$.
The states with $I \ge 53$ are
one dimensional, so the corresponding eigenvalues
(saturate quickly with $j$) may be  analytically derived \cite{Zhaoxx}.
These ``integer" eigenvalues
are best seen  in case of $J=J_{max}$ and becomes less
dominant for smaller $J$ and the same single-$j$ shell.

To understand the validity of the pair approximation and the
occurrence of these ``integer"
eigenvalues  we consider the pair basis of four nucleons
\begin{equation}
|j^4 [J_1 J_2]  I,M \rangle =
     \frac{1}{\sqrt{ N^{(I)}_{J_1 J_2; J_1 J_2}}}  \left(  A^{J_1\dag}  \times
A^{J_2\dag} \right)^{(I)}_M |0 \rangle,
\end{equation}
where $N^{(I)}_{J_1 J_2;J_1 J_2} $ is the diagonal matrtix element of the
normalization matrix
\begin{eqnarray}
N^{(I)}_{J^{\prime}_1 J^{\prime}_2; J_1 J_2} = \langle 0 |\left(  A^{J'_1}
\times A^{J'_2} \right)^{(I)}_M  \left(  A^{J_1\dag}  \times
A^{J_2\dag} \right)^{(I)}_M |0 \rangle   \nonumber   \\
=\delta_{J_1, J'_1} \delta_{J_2, J^{\prime}_2}
                    + (-)^I \delta_{J_1, J'_2} \delta_{J_2, J_1^{\prime}}
                    - 4 \hat{J}_1  \hat{J}_2  \hat{J^{\prime}_1}
\hat{J^{\prime}_2}
     \left\{%
     \begin{array}{ccc}
     j    & j  & J_1 \\
     j    & j  & J_2 \\
     J^{\prime}_1  & J^{\prime}_2 & I . \end{array}
     \right\}  .
\label{e5.12}
\end{eqnarray}
In general this basis is overcomplete and the normalization matrix
may have zero
eigenvalues for a given $I$.

The matrix elements of $H_J$ are \cite{Yoshinaga}
\begin{equation}
      \langle j^4 [J^{\prime}_1 J^{\prime}_2]I,M |
H_J  |j^4 [J_1 J_2] I,M\rangle
= -\frac {1}{{\sqrt{N^{(I)}_{J_1 J_2; J_1 J_2} N^{(I)}_{J^{\prime}_1
J^{\prime}_2;J^{\prime}_1 J^{\prime}_2 }} }}
\sum_{J^{\prime}={\rm even}}   N^{(I)}_{J_1 J_2; J J^{\prime}}
N^{(I)}_{J^{\prime}_1
J^{\prime}_2; J J^{\prime}}
     \label{matrix1}
\end{equation}
where $\hat{J}_1$ is a short hand notation of $\sqrt{2J_1+1}$.
There are two terms in   $N^{(I)}_{J_1 J_2; J J^{\prime}}$:
the second term is a nine-$j$ coefficient which is usually much less
than unity in
magnitude, in particular when $J$ is large and $I$ not large
(refer to Appendix A).
Neglecting this nine-$j$ symbol, the allowed states are $|j^4 [J_1 J_2]
I,M \rangle
\approx
      \left(  A^{J_1\dag}  \times
A^{J_2\dag} \right)^{(I)}_M |0 \rangle$  for $J_2 < J_1$, and $ |j^4
[J_1 J_1]  I,M
\rangle
\approx
     \frac {1}{\sqrt{2}} \left(  A^{J_1\dag}  \times
A^{J_1\dag} \right)^{(I)}_M |0 \rangle, I$ even only, and the
Hamiltonian matrix becomes
\begin{equation}
      \langle j^4 [J^{\prime}_1 J^{\prime}_2]I,M |
H_J  |j^4 [J_1 J_2] I,M\rangle
\approx  -\frac {(\delta_{J_1,J^{\prime}_1}\
\delta_{J_2,J^{\prime}_2} + (-1)^I
\ \delta_{J_1,J^{\prime}_2}\ \delta_{J_2,J^{\prime}_1})
}{\sqrt{1 + (-1)^I\delta_{J_1,J_2}}\ \sqrt{1 +
(-1)^I\delta_{J^{\prime}_1,J^{\prime}_2}}}\ (\delta_{J_1,J} + \delta_{J_2,J}).
     \label{matrixapprox}
\end{equation}
First of all we see that the matrix is diagonal which validates the pair
approximation.
Secondly the eigenvalues are either 0, $-1$, or $-2$ with their
corresponding wavefunctions having 0, 1 or 2 pairs with angular momentum $J$,
respectively. Therefore, the integer eigenvalues of many $I$ states
originate from both the special properties of
these nine-$j$ symbols and the validity of
$J$-pair truncation.

   From the  $J$-pair coupling scheme,
the number of $|j^4 [J_1 J_2] I,M\rangle$ with $J_1=J_2=2j-1$
is $\frac{1+ (-)^I }{2} $, and
the number of $|j^4 [J_1 J_2] I,M\rangle$ with $J_1=2j-1$, $J_2 < J_1$ and
$I < 2j-1$
is $[~ \frac{I}{2} ~]$
(the largest integer
not exceeding $\frac{I}{2}$). According to the above
discussion, the number
of states with eigenvalues $ \simeq -2$ is $\frac{1+ (-)^I }{2} $ and
the number of states with eigenvalues $ \simeq -1$ is $[~ \frac{I}{2} ~]$.
This is confirmed in {\it all} cases with $I < 2j-8$.
Eigenvalues not close to integers arise in states
with $2j-8 \le I \le 4j - 12$. These ``non-integer" eigenvalues
are found to be almost the same for the $2j-8 \le I \le 4j - 12$
states; an understanding of this regularity is in progress.

 From a more general expression of Eq. ~(\ref{matrix1}),
say, Eq. (5.8) of  \cite{NPSM},  we expect
that the ``integer" eigenvalues
appear not only in even systems but also appear in
odd-A systems. According to our numerical
results,   the pattern of ``integer" eigenvalues
also appear in the states with small $I$
for $j \ge 11/2$ and $n=3$,
and for $j \ge 23/2$ and $n=5$ etc.
For cases with $n=3$, a similar proof is readily obtained
in terms of 6-$j$ symbols. An explicit proof for  more nucleons
will be quite complicated.


\section {Exact Results for Angular Momentum Zero States}
We now come to the last point of this paper by pointing out that there
is only one non-zero eigenvalue for $I=0$ and $n=4$.
We define a new basis for the $I = 0$ states for the pair Hamiltonian
$H_J$
\begin{equation}
|j^4 J_1 \rangle= |j^4 [J_1J_1]I=M=0 \rangle
- \frac
{N^{(0)}_{J_1J_1;JJ}}{\sqrt{N^{(0)}_{JJ;JJ}N^{(0)}_{J_1J_1;J_1J_1}}}|j^4
[JJ]I=M=0
\rangle,\ (J_1 \neq J),
\label{basis}
\end{equation}
\begin{equation}
      |j^4 J \rangle= |j^4 [JJ]I=M=0 \rangle.
\end{equation}
The  unnormalized states $|j^4 J_1 \rangle, J_1 \neq J$, are
orthogonal with respect to  ${\it only}$   $  |j^4 [JJ]I=M=0 \rangle$.
   From  (\ref{matrix1}) one has
\begin{equation}
\langle j^4 [J^{\prime}_1J^{\prime}_1]0,0 | H_J  |j^4
[J_1J_1]0,0 \rangle =
- \frac
{N^{(0)}_{J_1J_1;JJ}N^{(0)}_{J_1J_1;JJ}}{\sqrt{N^{(0)}_{J^{\prime}_1J^{\prime}_1;JJ}
N^{(0)}_{J_1J_1;JJ}}},
\end{equation}
where $J^{\prime}_1,J_1=0, 2, \cdots, 2j-1$.
Using this formula, one easily confirms that all matrix elements of
the Hamiltonian in
the basis (\ref {basis}),
$ \langle j^4 J^{\prime}_1 | H_J  | j^4 J_1 \rangle,\
J^{\prime}_1 \neq J $, are
zero. Therefore all the eigenvalues for $n = 4$ and $I=0$ are zero
except for the single
state with both pairs having angular momentum $J$ and its eigenvalue is
$E_0^{J(j)}= - N^{(0)}_{JJ;JJ}$.

Since $H_J$ is a negative definite operator, its eigenvalues will be
negative or zero. From above we see that, for $I$ = 0, there is only one
state with a non-zero eigenvalue, $E_0^{J(j)}$. Therefore one expects
this eigenvalue to be the lowest in the spectrum because the eigenvalues
of $I \neq 0 $ states are more or less scattered in many states,
generally speaking.  Thus the probability that the $I=0^+$ is the lowest
state of four fermions in a single-$j$ shell in the presence of
random two-body interactions is expected to be
larger than the probability for all other angular momentum $I$,
according to the empirical rule of Ref. \cite{Zhaoxx}. For $j
\le 31/2$ there are only two exceptions ,
$j=7/2$ and $13/2$.

The sum  rule of diagonal matrix elements
\cite{Lawson} gives $\sum_J E_0^{J(j)}= -\frac{1}{2}
n(n-1) D_0^{(j)} = -6 D_0{(j)}$,
  where $D_0^{(j)}$ is the number of $I=0$ states and here $n=4$.
For $n$ = 4 the number of states is $D_0^{(j)}=
[(2j+3)/6]$ \cite{Ginocchio} which gives 1, 1, 1, 2, 2, 2, 3, 3, 3,
$\cdots$ for
$2j$=3, 5, 7, 9, 11, 13, 15, $\cdots$, etc., regularly.
Thus the staggering in the number of states is expected to be
  reflected in the staggering of the energy which was
pointed out in \cite{Zhaoxx,zelevinsky} but without an explanation.

\section{Summary}

In this paper we have shown that an attractive
$J$-th pairing  interactions  favors pairs
with angular momentum $J$ in  low-lying states of fermions in a
single-$j$ shell.  Therefore, one may use
pairs with angular momentum $J$ as building blocks of wavefunctions
of low-lying states. This is in contrast to repulsive pair interactions,
used, for example, in the fractional quantum Hall effect \cite
{lau,hald,wick}, for which the pair truncation approximation is not valid.

In addtion we discovered that the eigenvalues of states with low angular
momentum $I$ for pair Hamiltonians
with $J$ larger than $j$ are approximately integers.
We explain the origin of this
     fact for four nucleons in a single-$j$ shell from
     the validity of pair truncation and special properties
     of nine-$j$ symbols. We point out without details  that
the same holds for $n=3$ and $j \ge 11/2$ and for $n=5$ ($j\le 23/2$).


Finally we pointed out that there
is only one non-zero eigenvalue  for $I=0$ and $n=4$.  
This result, together with  the empirical rule of  ref. \cite{Zhaoxx}, 
provides a  simple  argument
for the large probability of angular momentum zero states to be the
lowest in energy for $n=4$.
Also for the first time, we demonstrated
that the staggering of this probability for $n=4$ vs. $j$
is directly related to the staggering of number of $I=0$ states vs. $j$.

\section{Acknowledgements}
One of the authors (ZYM) is grateful to
      the Japan Society for the Promotion of Science
     (contract No. P01021) for supporting his work, and  he also thanks
     Los Alamos National Laboratory for the warm hospitality extended to
him. This work
was partly supported by U.S. Department of Energy under contract
~W-7405-ENG-36.
\newpage

\section {Appendix A: Some nine-$j$ symbol formulae}

These formulae are obtained in the following two
steps: First, rewrite the nine-$j$ in terms of six-$j$ symbols, i.e.,
\begin{eqnarray}
    \left\{ \begin{array}{ccc}
    j    & j  & r_1 \\
    j    & j  & r_2 \\
    s_1  & s_2 & I  \end{array} \right\} = \sum_t
    (-)^{2t} (2t+1)
    \left\{ \begin{array}{ccc}
    j    & j  & r_1 \\
    r_1  & I & t  \end{array} \right\}
    \left\{ \begin{array}{ccc}
    j    & j  & r_2 \\
    j    & t  & s_2  \end{array} \right\}
    \left\{ \begin{array}{ccc}
    s_1    & s_2  & I \\
      t    &   j  & j  \end{array} \right\},    \nonumber
\end{eqnarray}
and second,  make use of the
analytical formulas of six-$j$.  Through
these examples (though we are unable to
get a universal formulae), one
sees that the nine-$j$ symbols in Eq. (3) are much less than unity
and may be neglected in (\ref{matrix1}) when $I$ is not very large.
\begin{eqnarray}
&&
    \left\{ \begin{array}{ccc}
    j    & j  & 2j-1 \\
    j    & j  & 2j-1\\
    2j-1  & 2j-1 & 0  \end{array} \right\} = -
    \frac{j (4j-3) \left[ (2j-1)! \right]^2}{(4j-1) (4j-1)!};
    \nonumber   \\
&&
    \left\{ \begin{array}{ccc}
    j    & j  & 2j-1 \\
    j    & j  & 2j-1\\
    2j-1  & 2j-1 & 2  \end{array} \right\} =
    \frac{j (8j^2 - 6j -3) \left[ (2j-1)! \right]^2}{(4j-1) (4j-3)(4j-1)!};
    \nonumber   \\
&&
    \left\{ \begin{array}{ccc}
    j    & j  & 2j-1 \\
    j    & j  & 2j-1\\
    2j-1  & 2j-1 & 4  \end{array} \right\} =
- \frac{3j(2j+1)(4j^2-3j-5) \left[ (2j-1)! \right]^2}{(4j-1) (4j-3)
(4j-5)(4j-1)!};
    \nonumber   \\
&&
    \left\{ \begin{array}{ccc}
    j    & j  & 2j-1 \\
    j    & j  & 2j-1\\
    2j-1  & 2j-1 & 6  \end{array} \right\} =
    \frac{5j(j+1)(2j+1)(8j^2-6j-21)
    \left[ (2j-1)! \right]^2}{(4j-1) (4j-3) (4j-5)(4j-7)(4j-1)!};
    \nonumber   \\
&&
    \left\{ \begin{array}{ccc}
    j    & j  & 2j-1 \\
    j    & j  & 2j-3\\
    2j-1  & 2j-3 & 2  \end{array} \right\} = -
    \frac{ j \left( 36 +j(4j-9) \left(19+2j(-9+4j) \right) \right)
     \left[ (2j-1)! \right]^2}{3(4j-3) (4j-5)(4j-1)!};
    \nonumber   \\
&&
    \left\{ \begin{array}{ccc}
    j    & j  & 2j-1 \\
    j    & j  & 2j-3\\
    2j-1  & 2j-3 & 3  \end{array} \right\} =
    \frac{ j (4j-3) \left(1+2j(4j-9) \right)
     \left[ (2j-1)! \right]^2}{6 (4j-5)(4j-1)!}.
    \nonumber
\end{eqnarray}

   From the above formulae one sees that
these nine-$j$ symbols are proportional to  $
    \frac{ \left[ (2j-1)! \right]^2}{(4j-1)!}$, and are
    very close to zero when $j$ becomes considerably large. For example,
    the  absolute values of these
    nine-$j$ symbols are less than $\sim 10^{-15}$ when $j=31/2$.

\section {Appendix B: A New Sum Rule for a six-$j$ symbol}

   From Eq. (3) one obtains
\begin{eqnarray}
&& N^{(0)}_{JJ;JJ} = 2+4(2J+1)
    \left\{ \begin{array}{ccc}
    j    & j  & J \\
    j    & j  & J  \end{array} \right\}. \nonumber
\end{eqnarray}
Since $-N^{(0)}_{JJ;JJ}$ is also the unique
eigenvalue of the $I=0$ eigenstate of $H_J$, one has a
sum rule that
\begin{eqnarray}
&& \sum_{{\rm even~}J} N^{(0)}_{JJ;JJ} = \frac{1}{2} n(n-1) D_0^{(j)}~,
\nonumber
\end{eqnarray}
where $n=4$, $D_0^{(j)}$ = $\left[~\frac{2j+3}{6}~\right]$ \cite{Ginocchio}.
One finally obtains that
\begin{eqnarray}
&& \sum_{{\rm even~}J} (2J+1)
    \left\{ \begin{array}{ccc}
    j    & j  & J \\
    j    & j  & J  \end{array} \right\} =
    \left\{ \begin{array}{cl}
    \frac{1}{2},    & {\rm if} ~ 2j=3k,    \\
    0,              & {\rm if} ~ 2j+2=3k,    \\
- \frac{1}{2},   & {\rm if} ~ 2j-2=3k.   \end{array} \right.
\end {eqnarray}

\begin{figure}

\caption{  Ground state angular momenta $I$ for four fermions in a
single-$j$ shell for the the pair Hamiltonian $H_J$ for
$J$=6 in (a) and 14 in (b) as a function of $j$.
The solid squares are ground state angular momenta obtained by truncating the
space of states to those with two pairs with
angular momentum $J$ only, and the open circles are ground state 
angular momenta
calculated by the diagonalizing the pair Hamiltonian in the full
shell model space.}
\end{figure}

\begin{figure}

\caption{A comparison of low-lying spectra calculated with wavefunctions
with  two pairs with
angular momentum
$J=14$ (the column on the left hand side)   and by a diagonalization
of the full
space (the column  in the middle  and the column on the right hand side) for
the case of four nucleons in a single-$j$ ($j=25/2$) shell.
The middle column plots the shell model states which are well reproduced
by the two $J=14$ pairs, and the right column plots
the shell model states which are not well reproduced by two $J=14$ pairs.
All the levels below $0^+_1$ in the full shell model
space  are included.
One sees that the low-lying states
with $I=2_1^+$, $6_1^+$, $12_1^+$, and $10_1^+$
are well reproduced. It is noted that the $0_1^+$ coupled by
two $J=14$ pairs is equivalent to that obtained by
a full shell model diagonalization; refer to the text.}

\end{figure}

\end{document}